\documentclass[prd, aps, showkeys, twocolumn, superscriptaddress, showpacs, nofootinbib, usenatbib, longbibliography]{revtex4-2}

\usepackage[
    colorlinks = true,
    linkcolor  = teal,
    citecolor  = teal,
    urlcolor   = teal,
    breaklinks = true
]{hyperref}
\usepackage{newtxtext,newtxmath}
\usepackage{graphicx}
\usepackage{amsmath}	
\graphicspath{{figures/}} 
\usepackage{color}

\usepackage{amssymb}
\usepackage[title]{appendix}
\usepackage[utf8]{inputenc}
\usepackage{subcaption}
\usepackage{changes} 

\definechangesauthor[name=Avi, color=blue]{avi}

\begin{document}
\title{{Bayesian power spectral density estimation for LISA noise based on penalized splines with a parametric boost}}
\author{Nazeela Aimen}
\affiliation{Department of Statistics, The University of Auckland, Auckland, New Zealand}
\author{Patricio Maturana-Russel}
\affiliation{Department of Statistics, The University of Auckland, Auckland, New Zealand}
\affiliation{Department of Mathematical Sciences, Auckland University of Technology, Auckland, New Zealand}
\author{Avi Vajpeyi}
\affiliation{Department of Statistics, The University of Auckland, Auckland, New Zealand}
\author{Nelson Christensen} 
\affiliation{Universit\'e C\^ote d'Azur, Observatoire de la C\^ote d'Azur, Artemis, CNRS, 06304 Nice, France}
\author{Renate Meyer}
\affiliation{Department of Statistics, The University of Auckland, Auckland, New Zealand}
 
\begin{abstract}
Flexible and accurate noise characterization is crucial for the precise estimation of gravitational-wave parameters.
We introduce a Bayesian method for estimating the power spectral density (PSD) of long, stationary time series, explicitly tailored for LISA data analysis.
Our approach models the PSD as the geometric mean of a parametric and a nonparametric component, combining the knowledge from parametric models with the flexibility to capture deviations from theoretical expectations.
The nonparametric component is expressed by a mixture of penalized B-splines. Adaptive, data-driven knot placement, performed once at initialization, removes the need for reversible-jump Markov chain Monte Carlo, while hierarchical roughness-penalty priors prevent overfitting.
Validation on simulated autoregressive AR(4) data demonstrates estimator consistency and shows that well-matched parametric components reduce the integrated absolute error compared to an uninformative baseline, requiring fewer spline knots to achieve comparable accuracy.
Applied to one year of simulated LISA X-channel (univariate) noise, our method achieves relative integrated absolute errors of $\mathcal{O}(10^{-2})$, making it suitable for iterative analysis pipelines and multi-year mission data sets.
\end{abstract}

\keywords{gravitational waves, PSD estimation, P-splines, LISA}


\maketitle

\section{Introduction}
The space-based Laser Interferometer Space Antenna (LISA) mission will open the low-frequency mHz band of gravitational wave astronomy, observing sources ranging from Galactic white-dwarf binaries to mergers of supermassive black holes~\cite{amaro-seoane_laser_2017}.
Unlike ground-based detectors, LISA will observe continuously for years (aside from occasional data gaps), with signals from multiple sources overlapping throughout the data stream.
This continuous observation eliminates the possibility of ``quiet'' periods for off-source noise estimation~\cite{littenberg_prototype_2023}, and LISA's unique noise characteristics—including instrumental noise, acceleration noise, and a dominant Galactic foreground—combined with multi-year observation times, demand computationally efficient methods capable of handling extremely long time series.
For context, a full LISA mission (4-6 years) sampled at 1 Hz would generate $\sim 10^8$ observations, making computational efficiency paramount.
In ground-based detectors such as LIGO–Virgo–KAGRA, noise is typically estimated from off-source segments using the Welch method~\cite{welch_use_1967}, averaging periodograms computed from multiple signal-free segments to obtain a consistent power spectral density (PSD) estimate~\cite{abbott_guide_2020}.
This PSD can be used as a plug-in estimate in the Whittle likelihood, enabling fast, low-latency analyses.
However, for LISA and other next-generation detectors such as the Einstein Telescope (ET)~\cite{punturo_einstein_2010} and Cosmic Explorer (CE)~\cite{reitze_cosmic_2019}, the absence of off-source, noise-only segments renders these approaches infeasible, motivating flexible, on-source PSD estimation methods that are both accurate and computationally efficient.

Accurate modeling of the detector PSD is critical for unbiased parameter estimation and well-calibrated uncertainties, since misestimating the noise spectrum can lead to biased parameter estimates and over-confident credible intervals~\cite{kirch_asymptotic_2024}.
Several algorithms have been developed for flexible PSD estimation, particularly for LISA.
BayesWave~\cite{cornish_bayeswave_2015} uses a reversible-jump Markov Chain Monte Carlo (RJMCMC)~\cite{Green:1995mxx,Umstatter:2005jd} algorithm to fit a trans-dimensional model with a Morlet-Gabor wavelet frame, simultaneously estimating the gravitational wave signal and the noise PSD via BayesLine~\cite{biscoveanu_quantifying_2020} using Lorentzians and splines.
Frequency-binned approaches such as SGWBinner~\cite{flauger_improved_2021} fit power laws in discrete bins, while Gaussian process models for the log-PSD~\cite{PhysRevD.109.083029} provide weakly parametric smoothing.
Several works model deviations from fixed parametric PSDs using B-splines or AKIMA splines with unknown numbers and locations of knots~\cite{muratore_impact_2023, santini_flexible_2025, baghi_uncovering_2023}, again relying on RJMCMC, and some incorporate parametric corrections to improve estimation of sharp spectral features.
Alternative Bayesian nonparametric methods~\cite{kirch_beyond_2019, edwards_bayesian_2019} avoid RJMCMC by inducing spline coefficients through a Dirichlet process prior on a cumulative distribution function. 
In particular, Kirch et al.~(2019)~\cite{kirch_beyond_2019} further improve the estimation of sharp PSD peaks by employing a nonparametrically corrected likelihood that combines autoregressive time-domain modeling with frequency-domain corrections.
While flexible, Markov Chain Monte Carlo (MCMC) based methods remain computationally expensive, particularly due to repeated transformations between time and frequency domains, highlighting the need for scalable alternatives for multi-year datasets.

Penalized splines (P-splines) provide a computationally efficient solution~\cite{maturana-russel_bayesian_2021, Eilers_Marx_Durbán_2015}.
By using a deliberately large B-spline basis with a hierarchical roughness penalty prior, P-splines allow flexible smoothing while controlling overfitting.

\newpage 

This work presents a novel log-P-spline algorithm tailored to LISA's computational demands. 
For LISA applications, we use knots that are uniformly spaced on a logarithmic scale.
This produces denser coverage at low frequencies, where LISA sensitivity is highest and spectral features such as $1/f$ acceleration noise, Galactic confusion noise, and instrumental artefacts are most prominent, while avoiding overparameterisation at high frequencies.
Logarithmic spacing is consistent with the standard log--log representation of spectral densities and ensures that the roughness penalty acts approximately uniformly across frequency decades.
This logarithmic parameterization is thus ideally suited to LISA's noise characteristics and scientific requirements.
Modeling the log-PSD removes the positivity constraint on the spline coefficients and allows the roughness penalty to operate uniformly across frequency decades.
Our approach uses a corrected likelihood exclusively in the frequency domain, avoiding the costly time-frequency transformations required by methods such as Kirch et al.~(2019)~\cite{kirch_beyond_2019}.
The PSD is modeled as a geometric mean of parametric and nonparametric components, allowing for flexible correction of potentially misspecified parametric models while retaining the speed and flexibility of P-splines.
The parametric component, incorporating known LISA characteristics such as acceleration noise scaling and instrumental transfer functions, further enhances the estimation of sharp spectral features.
The algorithm employs a blocked Whittle likelihood to handle extensive time series, dividing the dataset into segments with a shared PSD.
We demonstrate the method on a one-year-long simulated LISA time series sampled at 1 Hz ($\sim 3.15 \times 10^7$ observations), with computation times less than three minutes.

The paper is organized as follows.
Section~\ref{sec:model} details the P-spline methodology, including the blocked Whittle likelihood, penalty priors, and adaptive knot placement.
Section~\ref{sec:sim} demonstrates the benefits of incorporating well-fitting parametric models and provides empirical evidence of Bayesian spectral density estimate consistency.
Section~\ref{sec:LISAapp} applies the method to estimate the PSD of LISA instrumental $X$-channel noise, and Section~\ref{sec:Discussion} concludes with a discussion and outlook.

\section{P-spline Model for the Log PSD}\label{sec:model}

Let \( \mathbf{Z} \) be a time series of length \( n \), partitioned into \( J \) mean-centered segments \( Z^j = (Z_0^j, \dots, Z_{N-1}^j) \), for $j=1,\dots,J,$ of duration \( T \) and length \( N = T / \Delta_t \), where \( \Delta_t \) is the sampling interval (time between consecutive samples), sampled at frequency \( f_s = 1 / \Delta_t \) with Nyquist frequency \( f_{\mathrm{Ny}} = 1 / (2\Delta_t) \). 
The periodogram of each segment \( Z^j \) is given by
\begin{equation}
    I_j(f_l) = \frac{1}{N\Delta_f} \left| \sum_{t=0}^{N-1} Z^j_t e^{-i2\pi f_l t} \right|^2,
\end{equation}
where $f_l=l/(N\Delta_t)$ are the Fourier frequencies for $l = 0, \ldots, v,$ with $\,v= N/2-1$  when $N$ is even and $v=(N-1)/2$ when $N$ is odd, and $\Delta_f=1/(N\Delta_t)$. The periodograms have asymptotic independent exponential distributions with means equal to $TS_j(f_l)$, where $S_j(f_l)$ denotes the continuous two-sided spectral density matrix 
\begin{equation}
    S_j(f_l)=\frac{1}{2f_{Ny}}
\sum_{q=-\infty}^{\infty} \Gamma_j(q \Delta_t) \exp \left(-2\pi i f_l q\Delta_t \right),
\end{equation}
where $\Gamma_j(h)=\left( \gamma_{lm}(h)\right)=\mathbb{E}(Z^j_{t}Z^j_{t+h})$, with $h$ being the lag, is the auto-covariance function. 
This leads to the widely used Whittle likelihood function~\cite{whittle_curve_1957}, particularly prevalent in gravitational wave research~\cite{christensen_parameter_2022}. 
Under the additional assumptions of segment independence and identical PSDs $S(f_l) = S_j(f_l)$ for $j = 1, \ldots, J$, the blocked Whittle likelihood is obtained as
\begin{align}\label{llike}
\nonumber
    L(\textbf{Z}|S) & \propto \prod_{j=1}^J\prod_{l=1}^v\frac{1}{S(f_l)}e^{-I_j(f_l)/S(f_l)} 
    \\
     & \propto \exp \left\{ -J\sum_{l=1}^{v} \left(\log(S(f_l))+\frac{\bar{I}(f_l)}{S(f_l)}\right) \right\}, 
\end{align}
where 
\begin{align}\label{avgpdgrm}
\bar{I}(f_l)=\frac{1}{J}\sum_{j=1}^J I_j(f_l)
\end{align}
denotes the averaged periodogram.

To efficiently account for expected spectral features, we include a parametric PSD, \(S_\text{par}(f,\boldsymbol{\theta} _{\text{par}})\), representing these components of the spectrum within the likelihood. 
The full spectral density is then modeled as the geometric mean of this parametric component and a nonparametric P-spline component, \(S_\text{npar}(f,\boldsymbol{\theta} _{\text{npar}})\), allowing flexibility to capture any deviations while retaining the overall structure suggested by \(S_\text{par}(f,\boldsymbol{\theta} _{\text{par}})\). For simplicity, we refer to $S_\text{par}(f,\boldsymbol{\theta} _{\text{par}})$ and $S_\text{npar}(f,\boldsymbol{\theta} _{\text{npar}})$ as $S_\text{par}(f)$ and $S_\text{npar}(f)$, respectively. The full spectrum is therefore expressed as
\begin{align}\label{App:mod1}
        S(f_l)&=S_{\text{npar}}(f_l)^{1/2}S_\text{par}(f_l)^{1/2}\\\label{App:mod2}
        &=\left(\frac{S_\text{npar}(f_l)}{S_\text{par}(f_l)}\right)^{1/2} S_\text{par}(f_l)\\\label{App:mod3}
        &=c(f_l) S_\text{par}(f_l).
\end{align}
Equations (\ref{App:mod1}--\ref{App:mod3}) show that the geometric mean model turns out to be equivalent to modeling a correction of a parametric model as in Equation (13) of~\cite{muratore_impact_2023} where the parametric PSD is taken to be that of some instrumental design specification. 

The nonparametric correction  $c(f_l)$  adjusts for any variations that the parametric model fails to capture. The logarithm of the correction is modeled using the P-spline approach, i.e., by a linear combination of a large number $K$ of B-spline basis functions
\begin{equation} \log(c(f_l))=\sum_{k=1}^{K}\lambda_k b_{k,r}(f_l;\xi),
\end{equation}
where B-spline densities $b_{k,r}$ are normalized B-spline functions, i.e., integrate to $1$, of fixed degree $r$. $\boldsymbol{\xi}$ is the knot sequence $\boldsymbol{\xi}=\{f_{min}=\xi_0=\xi_1=\xi_r,\leq \xi_{r+1}\leq \dots \leq \xi_K=\xi_{K+1}=\dots=\xi_{K+r}=f_{max}\}$, which is fixed on the interval of frequencies ($f_{min}, f_{max}$) at which our periodogram is defined.
Here, $\boldsymbol{\lambda} = (\lambda_1,\dots,\lambda_K)^\top$ denotes the vector of spline coefficients, and $K$ is the total number of basis elements. The number of internal knots, which in our definition includes the boundary knots and is denoted $N_{\xi}$, is related to the number of basis functions by $K = N_{\xi} + r - 1$, where $r$ is the spline degree (e.g., $r=3$ for cubic B-splines).

Our method employs a B-spline basis defined on  logarithmically spaced knots for noise spectral density estimation of LISA. Logarithmic spacing is consistent with the standard visual and statistical treatment of spectral densities, which are frequently examined on a log–log scale.

The direct modeling of the logarithm of the PSD results in unconstrained values of $\lambda$, which stands in contrast to the approach in~\cite{maturana-russel_bayesian_2021}, where the weights were explicitly constrained to sum to one, thereby introducing an additional layer of complexity. The proposed formulation simplifies the MCMC sampler by removing this restriction and enhancing its computational efficiency.

B-spline functions require a predetermined set of knots. When their number is allowed to vary, methods such as RJMCMC or the stick-breaking representation of the Dirichlet process~\citep{edwards_bayesian_2019} become necessary.
Both approaches are challenging to tune, complicate the assessment of Markov chain convergence, and substantially increase computational cost. We select a large but fixed number of B-splines with predetermined knots in our proposal. Although this choice may naturally increase the risk of overfitting, it is mitigated by the penalty on the B-spline coefficients imposed through their prior distribution.

As shown in~\cite{maturana-russel_bayesian_2021}, judicious selection of the knot locations allows P-splines to retain the flexibility of B-splines while improving computational efficiency. 
We implement two options for placing the knots of the B-spline basis functions in our models. In the first approach, we place equidistant knots on the logarithmic scale, analogous to the traditional P-spline methods that employ equally spaced knots. 
In the second approach, the quantile-based knot scheme from~\cite{maturana-russel_bayesian_2021} is applied to the ratio of the averaged periodogram and the parametric PSD ($\bar{I}/S_\text{par}(f_l)$). This scheme allocates spline knots according to the empirical quantiles of this ratio, concentrating more knots in frequency regions where the periodogram exhibits strong deviations from the parametric model. 
Practically, the ratio is square-root transformed, standardized, and treated as a probability mass function whose cumulative distribution function is interpolated to define quantile locations. 
The resulting knot vector is fixed at initialization, so the computational cost of this step is negligible. 
This adaptive allocation enhances flexibility in regions where the parametric component underfits, while retaining smoothness elsewhere (for a more detailed discussion we refer the reader to~\cite{maturana-russel_bayesian_2021} and \cite{Ruppert2003}).
For clarity, throughout this work ``smoothness of the spectra'' refers to the smoothness of the ratio between the periodogram and the parametric model, since the P-spline prior is applied to the quantity $\log (c(f))$.

If the $\lambda$ coefficients were allowed to vary too freely, the resulting representation of the log correction would fluctuate excessively and lead to overfitting. Therefore, in the frequentist approach, one adds a penalty term to the likelihood function, e.g.,  $\phi( \Delta \lambda_k)=\phi (\lambda_k-\lambda_{k-1})$ or $\phi (\Delta^2\lambda_k)$ that penalizes the first or second order differences $\Delta^2\lambda_k= (\lambda_k-\lambda_{k-1})-(\lambda_{k-1}-\lambda_{k-2})$, respectively. 
Within the Bayesian context, instead of minimizing the penalized log-likelihood, the penalty term is naturally included in the prior distribution of the coefficients.
To avoid dependence on the choice of the penalty parameter $\phi$, 
a hierarchical prior structure is used as follows:
\begin{align}
    &\boldsymbol{\lambda} \mid \phi \sim \mathcal{N}_{K}(\mathbf{0},(\phi \mathbf{P})^{-1}), \\
    &\phi \mid \delta \sim \mathrm{Gamma}(\alpha_\phi, \delta\beta_\phi), \\
    &\delta \sim \mathrm{Gamma}(\alpha_\delta, \beta_\delta),
\end{align}
where $\mathbf{P}$ is a full-rank penalty matrix.
For non-equidistant knots, $\mathbf{P}$ is constructed using the derivative-based penalisation described by~\cite{wood_p-splines_2017} and used in~\cite{maturana-russel_bayesian_2021}.  
More explicitly, for a B-spline basis $\{b_{k,r}(f,\bm{\xi})\}_{k=1}^K$ of degree $r$ defined on an arbitrary knot sequence $\boldsymbol{\xi}$, the roughness penalty follows the standard Tikhonov regularisation framework and is given by the Gram matrix of the derivatives of the basis functions,
\begin{equation}
P_{kl}
  = \int_{f_{min}}^{f_{max}}
      b_{k,r}'(f; \bm{\xi})\, b_{l,r}'(f;\bm{\xi})\,\mathrm{d}f,
\qquad k,l = 1,\ldots,K .
\label{eq:penalty_matrix_general}
\end{equation}
This corresponds to a \emph{first-order} roughness penalty, i.e., penalisation of the squared $L^2$ norm of the first derivative of the spline $b_{k,r}'(f,\bm{\xi})$.
In our implementation, this matrix is evaluated directly by applying the linear differential operator to the B-spline basis, as provided by \texttt{skfda}~\cite{scikit-fda-repo, scikit-fda-paper}. 
The roughness penalty $\boldsymbol{\lambda}^\top \mathbf{P} \boldsymbol{\lambda}$ therefore discourages large local fluctuations in adjacent spline coefficients, enforcuing smoothness in the estimated spectrum.

The rate parameters $\beta_\phi$ and $\beta_\delta$ are given small prior values (e.g.\ $10^{-4}$), and the shape parameters $\alpha_\phi$ and $\alpha_\delta$ are set near~1 for robustness. 
These weakly informative Gamma priors are standard in hierarchical smoothing models, providing scale-invariant regularisation that encourages stable yet flexible behaviour across different datasets~\citep{JULLION2007,Bremhorst2013, maturana-russel_bayesian_2021}.
The hierarchical structure decouples the penalty strength from a fixed hyperparameter choice: the latent precision parameter $\phi$ governs the smoothness of the spline coefficients $\boldsymbol{\lambda}$, while the hyperparameter $\delta$ acts as a scaling prior that adaptively regularizes $\phi$. 
Larger $\phi$ values induce smoother spectra, whereas smaller $\phi$ values permit greater local flexibility. 
This formulation prevents over- or under-smoothing that could arise from arbitrarily chosen fixed penalty values.

Combining all unknown parameters into the parameter vector $\boldsymbol{\theta}=(\boldsymbol{\lambda}^\top,\phi,\delta)^\top$, the joint posterior distribution $p(\boldsymbol{\theta}|\textbf{Z})$ is given by
\begin{equation}
    p(\boldsymbol{\theta}|\textbf{Z})= L(\textbf{Z}|S)\times p(\boldsymbol{\lambda}|\phi,\delta)\times p(\phi|\delta)\times p(\delta) .
\end{equation}

We use a blocked Gibbs sampler to sample from the joint posterior $p(\boldsymbol{\theta}|\textbf{Y})$ by cyclically sampling from each of the full conditional posterior distributions.The parameters $\phi$ and $\delta$ can be sampled directly from their full conditional posterior distributions
\begin{align}
    \phi| \textbf{Z},\boldsymbol{\lambda},\delta\sim \mathrm{Gamma}\Big( \tfrac{K}{2}+\alpha_\phi, \tfrac{1}{2}\boldsymbol{\lambda}^\top\textbf{P}\boldsymbol{\lambda}+\delta\beta_\phi\Big),\label{eq:phi} \\
    \delta|\textbf{Z},\phi\sim \mathrm{Gamma}\Big(\alpha_\phi + \alpha_\delta, \beta_\phi\phi+\beta_\delta\Big)\label{eq:delta} .
\end{align}

Within the Gibbs-sampling framework, $\boldsymbol{\lambda}$ is sampled either using a Metropolis–Hastings method for component-wise updates \citep{maturana-russel_bayesian_2021} or an adaptive Metropolis–Hastings (AMH) algorithm for joint updates \citep{roberts_examples_2009}. 

For computational efficiency, the knots are placed at the beginning of the algorithm and remain fixed throughout. To ensure sufficient flexibility, we can use a large $N_{\xi}$, as illustrated in the Application section, so that the model can capture the underlying variations of the PSD while the penalty prior guards against overfitting.

\section{Simulation study}\label{sec:sim}
\begin{figure}[htbp]
    \centering
    \begin{subfigure}[t]{0.95\columnwidth}
        \centering
        \includegraphics[width=\columnwidth]{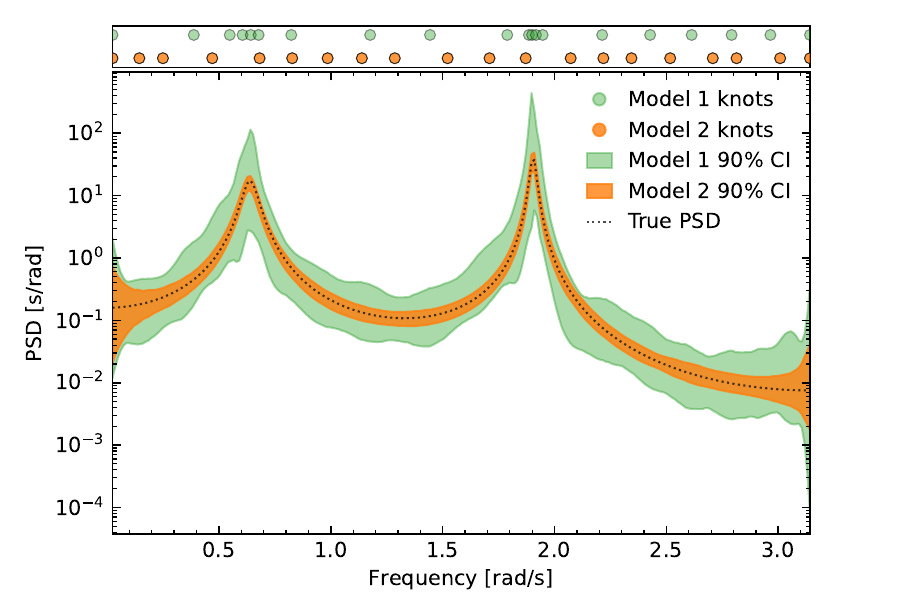}
         \caption{PSD estimates for one $n=512$ realization under WN and AR(4) parametric models. Shaded regions show $90\%$ credible intervals; dots mark spline knots.}
        \label{fig:PSD}
    \end{subfigure}

    \smallskip
    
    \begin{subfigure}[t]{0.95\columnwidth}
        \centering
        \includegraphics[width=\columnwidth]{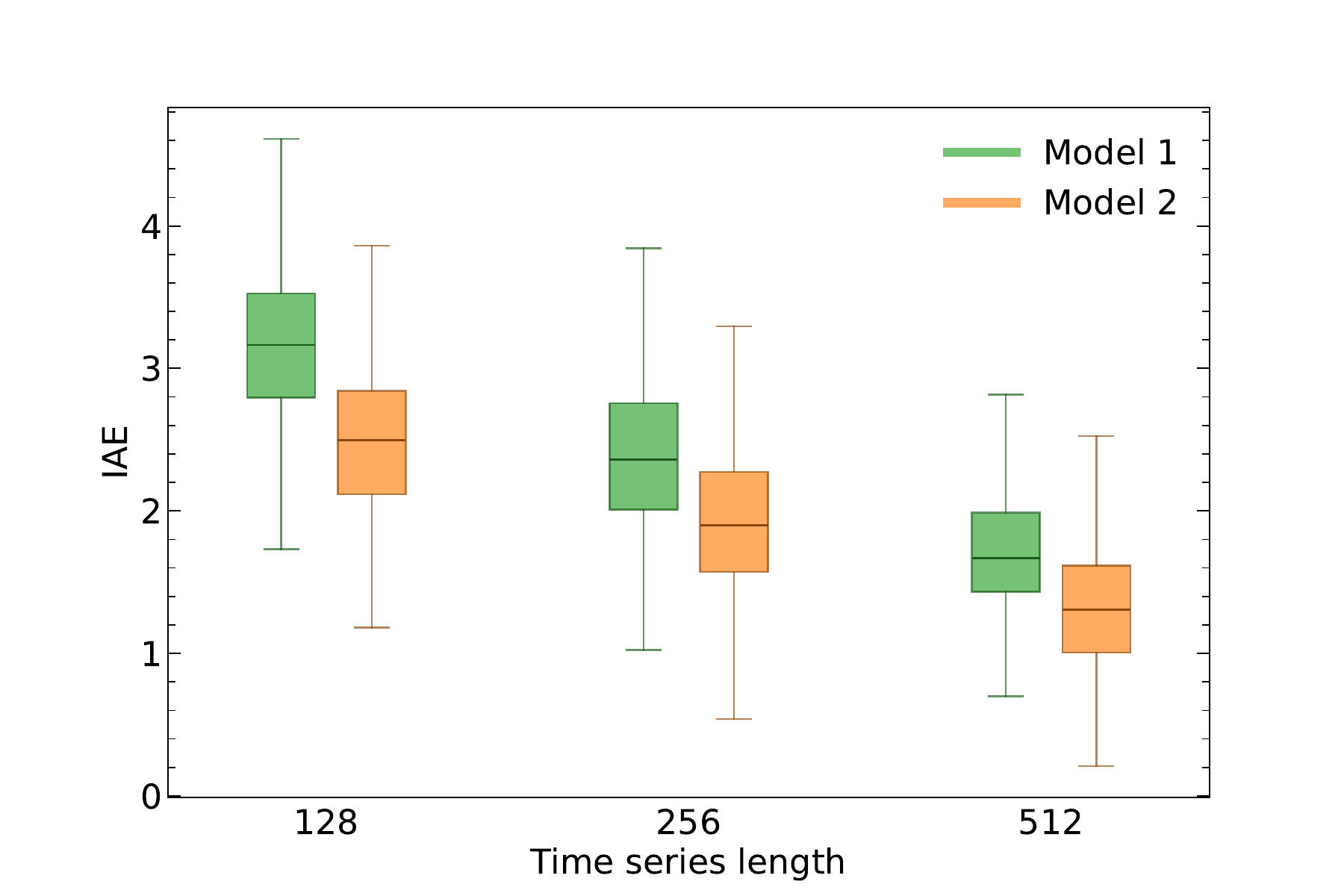}
        \caption{IAE distributions for different $n$.        Boxes indicate the interquartile range, whiskers the full range, and horizontal lines the median.}
        \label{fig:boxplot}
    \end{subfigure}

    \smallskip

    \begin{subfigure}[t]{0.95\columnwidth}
        \centering
        \includegraphics[width=\columnwidth]{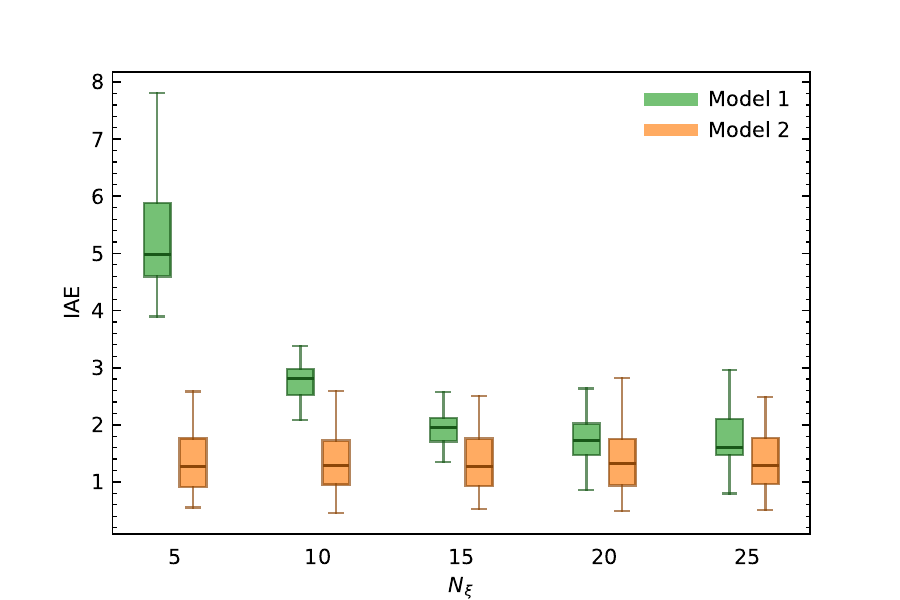}
        \caption{IAE distributions for different numbers of knots for $n=512$. 
        Boxes indicate the interquartile range, whiskers the full range, and horizontal lines the median.}
        \label{fig:iae_vs_knots}
    \end{subfigure}
    
    \caption{Performance comparison between semi-parametric PSD estimators. 
    (a) PSD estimates for a single realization; 
    (b) IAE distributions across dataset sizes; 
    (c) Sensitivity of IAE to the number of knots. 
    Model~1 (green) and Model~2 (orange) are shown throughout.
    }
    \label{fig:ar4_vs_ar0}
\end{figure}

We evaluate the accuracy of our method by simulating stationary AR(4) time series and estimating their PSDs. 
The theoretical AR(4) PSD is given by
\begin{equation}
S_{\text{AR}}(f) = \frac{\sigma^2}{2\pi}\frac{1}{\left|1-\sum_{k=1}^{4}a_k \exp(-2\pi i f k)\right|^2},
\end{equation}
where $\sigma^2$ is the variance and $(a_1,a_2,a_3,a_4)$ are the AR model parameters. 
We set $f_s = 1$, the variance to unity, and $a_1 = 0.9$, $a_2 = -0.9$, $a_3 = 0.9$, $a_4 = -0.9$, following previous work~\cite{edwards_bayesian_2019, maturana-russel_bayesian_2021}. 
These parameters produce a spectrum with two sharp peaks (Figure~\ref{fig:PSD}).
Estimation accuracy is measured using the integrated absolute error,
\begin{equation}\label{eq:iae}
\text{IAE} = \int_{f_{\min}}^{f_{\max}} |\hat{S}(f) - S(f)| \text{d}f,
\end{equation}
where $\hat{S}(f)$ is the pointwise posterior median.

We simulate series of length $n \in \{128, 256, 512\}$, generating $500$ independent realizations for each $n$. 
For each realization, we estimate the log-spectral density by running $20,000$ iterations with a $1,000$-iteration burn-in, and a thinning factor of $10$, resulting in $1,900$ samples for posterior inference. 
We calculate the covariance matrix from a large posterior sample to initialize the AMH algorithm across the simulations to generate the posterior samples.

We compare the two following semi-parametric PSD models within the framework of Section~\ref{sec:model}:
\begin{itemize}
    \item Model 1: $S_\text{par}(f_l)$ is a flat, uninformative white-noise (WN) PSD, with deviations absorbed by the spline correction, $c(f_l)S_\text{WN}(f_l)$.
    \item Model 2: $S_\text{par}(f_l)$ is an AR(4) PSD with parameters estimated by likelihood maximization, corrected in the same way, $c(f_l)S_{\text{AR}(4)}(f_l)$.
\end{itemize}

For both models, we place $20$ knots using the quantile-based method applied to the ratio of the periodogram and the parametric PSD. 
Using the same number of knots ensures a fair comparison between the models, isolating differences due to the parametric component rather than spline resolution.

To quantify the contribution of the initial parametric component before any spline correction, we calculate the median IAE between $S_\text{par}(f_l)$ and the true spectrum.
The $S_\text{WN}(f_l)$ provides a poor fit, with median IAE values consistently slightly above $5$ across sample sizes.
In contrast, $S_\text{AR(4)}(f_l)$ already captures much of the spectral structure, producing substantially lower errors that decrease from roughly $2.5$ at $n=128$ to $1.3$ at $n=512$.
Hence, the choice of $S_\text{par}(f)$ directly affects how much structure remains for the splines to recover.

Figure~\ref{fig:PSD} shows the results for one realization with $n=512$: the spline alone (Model 1) recovers the peaks with an IAE of $0.54$. In contrast, the AR(4)-assisted model (Model 2) requires only minor local adjustments, achieving an IAE of $0.21$.
The credible intervals are also narrower in Model~2, with a median width of $0.09\mathrm{s/rad}$ compared to $0.37\mathrm{s/rad}$ in Model~1. 
However, this difference is partly due to the conditioning on fixed AR parameters. The true credible intervals in Model~2  would be wider if parametric uncertainty were fully propagated, for instance, by jointly sampling spline and AR coefficients instead of conditioning on fixed AR estimates. However, sampling both spline and AR coefficients would be problematic due to degeneracies in the likelihood. In many applications, the parametric template will not be estimated from the data but will be prespecified.
For instance, a parametric template in the LISA space mission incorporates known noise components: thermal, charging, and magnetic effects causing low-frequency test mass acceleration noise, plus optical metrology noise dominating at high frequencies.


Figure~\ref{fig:boxplot} summarizes performance across all $n$. 
The IAE decreases with sample size for both models, but Model 2 consistently outperforms Model 1 with the median differences of IAEs reducing from $0.67$ to $0.36$ with increasing length of the time series. 
Thus, an informative $S_\text{par}(f)$ reduces both bias and variance, while the spline ensures robustness against parametric misspecification. 
%
These results underscore the importance of selecting a parametric model that accurately captures the variations of the PSD, even though its influence on the analysis diminishes as the sample size grows.

To assess the impact of $N_{\xi}$ on estimation accuracy, we analyze 50 simulated AR(4) datasets of length $n = 512$ and estimate the PSD using Models~1 and 2 with $N_{\xi}=\{5, 10, 15, 20, 25 \}$. 
Figure~\ref{fig:iae_vs_knots} displays the IAE as a function of  $N_{\xi}$ for both models. 
For Model~2, the IAE remains largely unchanged across the tested range, indicating that the hierarchical roughness-penalty prior effectively guards against overfitting even when the parametric component captures nearly all spectral structure. 
In contrast, Model~1 shows a reduction in IAE as $N_{\xi}$ increases from $5$ to approximately $10$, beyond which additional knots yield only marginal improvements. 
These results demonstrate that a well-specified parametric model reduces the number of knots needed for the P-spline correction to achieve a given accuracy.
When no suitable parametric approximation is available, more knots are required, but the non-parametric correction still performs well.


\section{Application to LISA Noise}\label{sec:LISAapp}
We demonstrate the application of the blocked semi-parametric P-spline framework to second-generation TDI-$X$ channel data dominated by low-frequency test-mass acceleration (TM) and high-frequency optical metrology readout (OMS) noise. We analyze simulated LISA noise timeseries from the public Zenodo repository (DOI: \href{https://doi.org/10.5281/zenodo.15698080}{10.5281/zenodo.15698080}), generated using \texttt{LISA Instrument v1.1.1}~\cite{bayle_lisa_2025,bayle_unified_2023}.

\subsection{Noise Models}

The simulation implements LDC Spritz noise curves under simplified orbital dynamics with fixed 8.3-second light travel times for inter-spacecraft laser beam propagation. 
Second-generation Michelson TDI combinations are computed via~\texttt{PyTDI}~\cite{staab_pytdi_2025,tinto_second-generation_2023}. 
The simulated noise PSD $S_{\rm X}(f)$ can be approximated using analytical models for test-mass acceleration noise~\cite{bayle_unified_2023}
\begin{equation} 
        S_{\rm TM}(f)=a^2_{\rm TM}\Big[ 1+\Big(\frac{f_1}{f}\Big)^2 \Big]\Big[1+\Big(\frac{f}{f_2}\Big)^4\Big]
        \Big(\frac{1}{2\pi fc}\Big)^2,
\end{equation}
where 
$a_{\rm TM}=2.4\times 10^{-15}\mathrm{ms}^{-2}$, 
$f_1=4\times 10^{-4}\,\mathrm{\mathrm{Hz}}$, 
$f_2=8\times 10^{-3}\,\mathrm{\mathrm{Hz}}$, 
and readout noise~\cite{bayle_unified_2023}

\begin{equation} 
    \label{eq:oms}
    S_{\rm OMS}(f)=a^2_{\rm OMS}\Big[ 1+\Big(\frac{f_3}{f}\Big)^4 \Big]\Big(\frac{2\pi f }{c}\Big)^2,
\end{equation}
where $a_{\rm OMS}=7.9\times 10^{-12}\mathrm{ms}^{-2}$, $f_3=2\times 10^{-3}\,\mathrm{Hz}$. 
The TM and OMS noise components are given in units of $\text{Hz}$ (see the Zenodo data release for more details on the data generation~\cite{bayle_lisa_2025}).
Taking into account the transfer functions through TDI~\cite{quang_nam_time-delay_2023}, the theoretical approximate PSD for the $X$ channel is given by
\begin{align}\label{eq:transfer_func}
    \nonumber S_{\text{X}}(f)=&16\sin^2(2\pi fL/c)\sin^2(4\pi fL/c)\\
    &(4S_{\rm OMS}(f)+[3+\cos(4\pi fL/c)]S_{\rm TM}(f)).
\end{align}

\subsection{Application of P-spline framework}

\begin{figure*}[ht]
    \centering
    \includegraphics[width=1\linewidth]{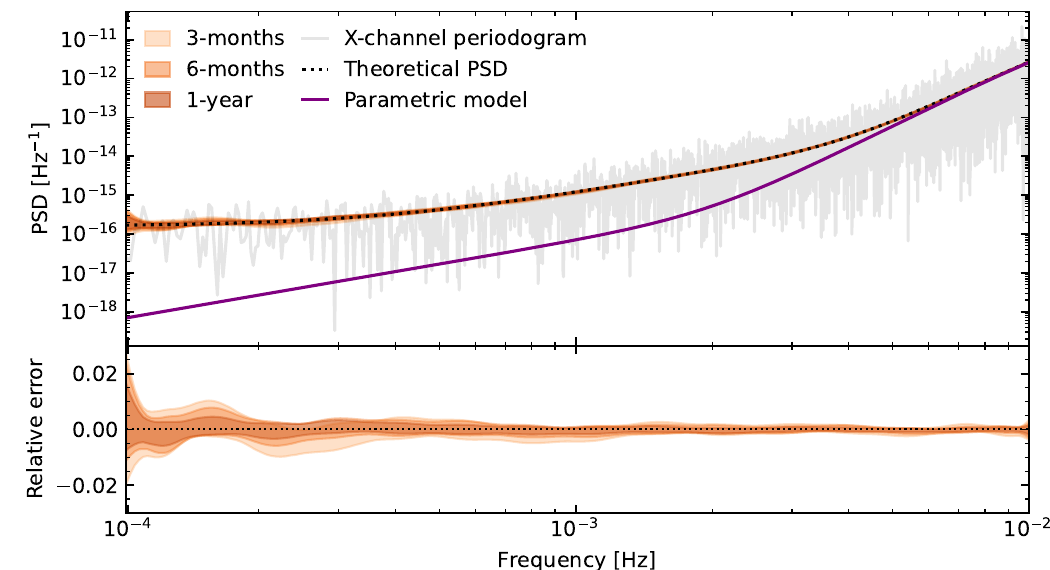}
    \caption{Power-spectral-density (PSD) estimation for the second-generation Michelson TDI $X$ channel using the OMS parametric noise model. Results are shown for data sets of 3\,months, 6\,months, and 1\,year (light-to-dark red). 
    \textit{Top:} shaded regions show the middle $90\%$ credible interval (CI) for each observing time. The dashed black curve is the theoretical PSD $S_{\rm X}$, solid purple curve is the OMS parametric PSD, and the gray trace is a representative 5-day block periodogram. 
    \textit{Bottom:} relative error of the CI of log PSD against $\log(S_{\rm X})$. Longer observation times yield narrower CIs and smaller relative errors.}
    \label{fig:psd-length-comparison}
\end{figure*}

We partition LISA noise time series of duration $T \in \{3, 6, 12\}$ months into $J \in \{18, 36, 73\}$ five-day segments and apply a Kaiser window with $\beta=5$ (following noise-4a dataset settings~\cite{bayle_lisa_2025}) to each segment. 
We then apply the blocked likelihood with 15 equidistant logarithmically-spaced frequency knots to the segmented periodogram.

The semi-parametric model employs the OMS noise with the transfer function, $16\sin^2(2\pi fL/c)\sin^2(4\pi fL/c)4S_{\rm OMS}(f)$ as a parametric model.
At low frequencies, the $X$-channel PSD is dominated by test-mass acceleration noise, whereas at high frequencies, the OMS component governs the spectral behavior. 
To emulate conditions where the true PSD may diverge from pre-flight expectations owing to environmental or instrumental effects, we adopt a deliberate, incomplete parametric model (OMS only), leaving the nonparametric component to capture the resulting discrepancies.
We generate 7,000 posterior samples after 15,000 iterations with 1,000 burn-in and a thinning factor of 2.

Figure~\ref{fig:psd-length-comparison} compares PSD estimates for three observing durations (3, 6, and 12 months).
The top panel displays  $90\%$ uniform credible intervals (CIs) for each duration, with darker orange shading corresponding to longer observing periods.
The dashed black curve displays $S_{\rm X}(f)$, while the gray trace shows a representative 5-day block periodogram. The purple curve shows our parametric model (OMS parametric PSD).
As observing time increases, the CIs contract across the LISA band, reflecting the reduced uncertainty from additional data. 
The posterior median closely tracks $S_{\rm X}(f)$ despite the misspecified parametric component, demonstrating that the P-splines effectively correct for the parametric model inadequacies.
Moreover, the analysis yields precise estimates in the central millihertz region where LISA exhibits maximum sensitivity to a stochastic gravitational wave background~\cite{Christensen_2018,babak_lisa_2021}.

The bottom panel quantifies these improvements through relative errors of CI of log PSD with respect to $\log(S_{\rm X}(f))$. 
For each duration, we plot $(q(f)-\log(S_{\rm X}(f)))/|\log(S_{\rm X}(f))|$, where $q(f)$ are the uniform $5\%$ and $95\%$ credible interval bounds of log PSD, respectively.
The relative errors decrease with longer observing periods and are largest at lower frequencies due to reduced spectral resolution in that regime.

\begin{table}[htbp]
\centering
\small
\caption{Median relative integrated absolute error (RIAE), 
runtime and median bulk effective sample size (ESS) of the spline coefficients for LISA Spritz noise.}
\label{tablelisa}
\begin{tabular}{lccc}
 & 3\,months & 6\,months & 1\,year \\
\hline
RIAE ($\times 10^{-2}$) & $1.47$ & $0.52$ & $0.25$ \\
Runtime (minutes) & 2.68 & 2.46 & 2.29 \\
Median bulk ESS & 426.88 & 451 & 507 \\
\end{tabular}
\end{table}
The relative integrated absolute error (RIAE) is calculated using the following formula
\begin{equation}
    \text{RIAE}= 
    \frac{\int_{f_{\min}}^{f_{\max}} |\hat{S}(f) - S_\text{X}(f)|\text{d}f}{\int_{f_{\min}}^{f_{\max}}S_\text{X}(f)\text{d}f},
\end{equation} 
with $\hat{S}(f)$ being the pointwise posterior median. It remains $\mathcal{O}(10^{-2})$ across all durations (Table~\ref{tablelisa}), demonstrating robust PSD reconstruction. 
Computational efficiency is maintained at $\sim 2.5$ minutes per analysis, independent of $T$, since the Whittle likelihood operates on a fixed-length averaged periodogram (Equation~\eqref{avgpdgrm}), with fixed knot locations. 
The effective sample size exhibits modest increases with additional data, indicating marginal improvements in sampling efficiency.
These results demonstrate our semi-parametric framework's ability to recover the PSD accurately despite deliberate parametric misspecification at low frequencies.  

The approach consistently yields minimal uncertainties in the central millihertz band—precisely where stochastic gravitational wave background detection sensitivity is maximized~\cite{robson_construction_2019}. This performance remains robust across all timescales tested, supporting the framework's applicability for LISA noise characterization across mission-relevant timescales.

\section{Discussion}\label{sec:Discussion}

We have developed a Bayesian semiparametric method for estimating the PSD of a stationary time series, focusing on LISA noise characterization.   
The PSD is expressed as the geometric mean of a parametric model and a nonparametric P-spline correction, combining the efficiency of a well-specified parametric form with the flexibility and robustness of a nonparametric approach.  
Non-uniform knot placement focuses model complexity on regions where the data are most informative.
A hierarchical penalty prior regularizes the spline coefficients, mitigating overfitting without compromising adaptability.

Simulation results demonstrate that the method is both accurate and consistent.  
When the parametric component is well-specified, as in the AR(4) case, the nonparametric correction only applies minor, targeted adjustments, resulting in substantially lower IAE than for a flat-spectrum starting model.  
In both cases, the IAE decreases with increasing sample size, illustrating that the estimator converges to the true PSD in the large-sample limit. Moreover, improved parametric templates led to markedly better computational efficiency, with fewer knots achieving accurate estimates.  
Therefore, the ability to achieve high accuracy with relatively short time series highlights the value of incorporating informed parametric structure when available, as it also reduces the computational effort required to attain comparable accuracy when a suitable parametric model is unavailable or does not provide a good fit.

For long LISA-like time series, the blocked Whittle likelihood combined with the P-spline correction produced precise PSD estimates at low computational cost.  
Across data lengths from $3$ to $12$ months, median IAE remained below \( 1.5\times 10^{-2} \), and runtimes were stable at around three minutes due to the use of averaged periodograms.  


The method has several practical advantages for LISA analysis.  
It can accommodate partial parametric knowledge without requiring the full PSD to be specified a priori.  
It remains computationally feasible for year-long data sets, enabling repeated use in iterative global-fit analysis pipelines~\cite{Littenberg:2023xpl,Katz:2024oqg,Deng:2025wgk}.  
Moreover, its nonparametric correction can capture subtle deviations from nominal instrument models, making it a valuable diagnostic tool.


There are also natural extensions.  
GPU acceleration should substantially reduce runtimes, making the method suitable for large-scale simulations and Monte Carlo studies.  
Extending the framework to separate stochastic gravitational wave backgrounds from instrument noise~\cite{Boileau:2021sni} would enable joint noise and signal inference.  
Generalising the spline correction to cross-spectral density matrices would allow simultaneous modelling of the correlated \( X \), \( Y \), and \( Z \) channels, providing a more complete representation of the LISA data.
Beyond LISA, the method could be adapted to estimate PSDs in ground-based detectors such as those in the LIGO-Virgo-KAGRA network, where it may assist in rapid noise modelling and continuous-wave searches.
Using spline correction for cross-spectral density matrices could also be useful if correlated noise is present in the Einstein Telescope triangular configuration~\cite{Janssens:2022xmo}.
An additional avenue is developing a time–frequency formulation that enables the estimation of evolving PSDs in the presence of non-stationary noise, which is relevant for both LISA and terrestrial detectors.

In conclusion, the proposed framework offers a fast, accurate, and adaptable tool for PSD estimation in high-precision astrophysical applications and beyond.  
Its ability to leverage partial parametric knowledge while retaining nonparametric flexibility makes it well-suited for the complex and evolving noise environment expected in LISA.  
The method is compatible with existing global-fit approaches, allowing it to serve as a dedicated noise-modelling component within broader inference pipelines for LISA data analysis.
These properties make it a practical addition to the LISA data-analysis toolkit and a generally applicable approach for other long-baseline spectral estimation problems.



\subsection*{Data and Software Availability}
The software developed for this project is open-source and publicly available from the GitHub repository \url{https://github.com/nz-gravity/npc.git}, which contains all source code, example scripts, and configuration files needed to reproduce the results. Installation instructions and dependency information (tested on Python~$\geq$3.10$)$ are provided in the repository \texttt{README}. 
The software is released under the MIT License, permitting free use, modification, and distribution. We also plan to release a PyPI package, fully compatible with JAX and GPU acceleration, allowing for easy installation and efficient execution on modern hardware. 
The datasets generated are available available at the public Zenodo repository~\cite{aimen_bayesian_2025}. LISA time series data is available at~\cite{bayle_lisa_2025}.

\begin{acknowledgments}
We thank Quentin Baghi, Jean-Baptiste Bayle, Matt Edwards, Martina Muratore, and others in the LISA Noise Non-Stationarities Group (part of the ``Deep analysis group'' of the Distributed Data Processing Centre, DDPC) for helpful discussions.
We additionally thank Jean-Baptiste Bayle for providing the LISA dataset, and the University of Glasgow for the computing resources that supported JB in simulating the dataset.
NA, NC, PMR, RM, and AV gratefully acknowledge support from the Marsden Fund Council grant MFP-UOA2131, funded by the New Zealand Government and managed by the Royal Society Te Apārangi.  
For this project, NC has received financial support from the CNRS through the MITI interdisciplinary programs, and the French Agence Nationale de la Recherche.           
This work was performed on the OzSTAR national facility at Swinburne University of Technology. The OzSTAR program receives funding in part from the Astronomy National Collaborative Research Infrastructure Strategy (NCRIS) allocation provided by the Australian Government, and from the Victorian Higher Education State Investment Fund (VHESIF) provided by the Victorian Government.  
\end{acknowledgments}

\bibliographystyle{unsrt}
\bibliography{new_ref_test}

@article{JULLION2007,
title = {Robust specification of the roughness penalty prior distribution in spatially adaptive Bayesian P-splines models},
journal = {Computational Statistics \& Data Analysis},
volume = {51},
number = {5},
pages = {2542-2558},
year = {2007},
issn = {0167-9473},
doi = {https://doi.org/10.1016/j.csda.2006.09.027},
url = {https://www.sciencedirect.com/science/article/pii/S0167947306003549},
author = {Astrid Jullion and Philippe Lambert},
}

@ARTICLE{Bremhorst2013,
       author = {{Bremhorst}, Vincent and {Lambert}, Philippe},
        title = "{Flexible estimation in cure survival models using Bayesian P-splines}",
      journal = {arXiv e-prints},
         year = 2013,
        month = dec,
          eid = {arXiv:1312.2369},
        pages = {arXiv:1312.2369},
          doi = {10.48550/arXiv.1312.2369},
archivePrefix = {arXiv},
       eprint = {1312.2369},
 primaryClass = {stat.ME},
       adsurl = {https://ui.adsabs.harvard.edu/abs/2013arXiv1312.2369B}
}

@article{santini_flexible_2025,
    author = "Santini, Alessandro and Muratore, Martina and Gair, Jonathan and Hartwig, Olaf",
    title = "{A flexible, GPU-accelerated approach for the joint characterization of LISA instrumental noise and Stochastic Gravitational Wave Backgrounds}",
    eprint = "2507.06300",
    note={arXiv:2507.06300 [gr-qc]},
    archivePrefix = "arXiv",
    primaryClass = "gr-qc",
    month = "7",
    year = "2025"
}

@article{reitze_cosmic_2019,
    author = "Reitze, David and others",
    title = "{Cosmic Explorer: The U.S. Contribution to Gravitational-Wave Astronomy beyond LIGO}",
    eprint = "1907.04833",
    archivePrefix = "arXiv",
    primaryClass = "astro-ph.IM",
    reportNumber = "LIGO-P1900316",
    journal = "Bulletin of the American Astronomical Society",
    volume = "51",
    number = "7",
    pages = "035",
    year = "2019"
}

@misc{kirch_asymptotic_2024,
	title        = {Asymptotic considerations in a {Bayesian} linear model with nonparametrically modelled time series innovations},
	author       = {Kirch, Claudia and Meier, Alexander and Meyer, Renate and Tang, Yifu},
	year         = 2024,
	publisher    = {arXiv},
	doi          = {10.48550/ARXIV.2409.16207},
	url          = {https://arxiv.org/abs/2409.16207},
	urldate      = {2025-09-17},
	copyright    = {Creative Commons Attribution 4.0 International},
    note         = {arXiv:2409.16207 [math.ST]}
}

@misc{babak_lisa_2021,
	title        = {{LISA} {Sensitivity} and {SNR} {Calculations}},
	author       = {Babak, Stanislav and Hewitson, Martin and Petiteau, Antoine},
	year         = 2021,
	publisher    = {arXiv},
	doi          = {10.48550/ARXIV.2108.01167},
	url          = {https://arxiv.org/abs/2108.01167},
	urldate      = {2025-09-17},
    note         = {arXiv:2108.01167 [astro-ph.IM]},
	copyright    = {Creative Commons Attribution Non Commercial Share Alike 4.0 International}
}

@article{muratore_impact_2023,
    author = "Muratore, Martina and Gair, Jonathan and Speri, Lorenzo",
    title = "{Impact of the noise knowledge uncertainty for the science exploitation of cosmological and astrophysical stochastic gravitational wave background with LISA}",
    eprint = "2308.01056",
    archivePrefix = "arXiv",
    primaryClass = "gr-qc",
    doi = "10.1103/PhysRevD.109.042001",
    journal = "Physical Review D",
    volume = "109",
    number = "4",
    pages = "042001",
    year = "2024"
}

@article{PhysRevD.109.083029,
  title = {Weakly parametric approach to stochastic background inference in LISA},
  author = {Pozzoli, Federico and Buscicchio, Riccardo and Moore, Christopher J. and Haardt, Francesco and Sesana, Alberto},
  journal = {Physical Review D},
  volume = {109},
  issue = {8},
  pages = {083029},
  numpages = {15},
  year = {2024},
  month = {Apr},
  publisher = {American Physical Society},
  doi = {10.1103/PhysRevD.109.083029},
  url = {https://link.aps.org/doi/10.1103/PhysRevD.109.083029}
}

@article{robson_construction_2019,
	title        = {The construction and use of {LISA} sensitivity curves},
	author       = {Robson, Travis and Cornish, Neil J. and Liu, Chang},
	year         = 2019,
	month        = may,
	journal      = {Classical and Quantum Gravity},
	volume       = 36,
	number       = 10,
	pages        = 105011,
	doi          = {10.1088/1361-6382/ab1101},
	issn         = {0264-9381, 1361-6382},
	url          = {https://iopscience.iop.org/article/10.1088/1361-6382/ab1101},
	urldate      = {2025-09-17}
}

@article{tinto_second-generation_2023,
	title        = {Second-generation time-delay interferometry},
	author       = {Tinto, Massimo and Dhurandhar, Sanjeev and Malakar, Dishari},
	year         = 2023,
	month        = apr,
	journal      = {Physical Review D},
	volume       = 107,
	number       = 8,
	pages        = {082001},
	doi          = {10.1103/PhysRevD.107.082001},
	issn         = {2470-0010, 2470-0029},
	url          = {https://link.aps.org/doi/10.1103/PhysRevD.107.082001},
	urldate      = {2025-09-17},
	language     = {en}
}

@article{edwards_bayesian_2019,
	title        = {Bayesian nonparametric spectral density estimation using {B}-spline priors},
	author       = {Edwards, Matthew C. and Meyer, Renate and Christensen, Nelson},
	year         = 2019,
	month        = jan,
	journal      = {Statistics and Computing},
	volume       = 29,
	number       = 1,
	pages        = {67--78},
	doi          = {10.1007/s11222-017-9796-9},
	issn         = {0960-3174, 1573-1375},
	url          = {http://link.springer.com/10.1007/s11222-017-9796-9},
	urldate      = {2025-09-17},
	language     = {en}
}

@article{kirch_beyond_2019,
	title        = {Beyond {Whittle}: {Nonparametric} {Correction} of a {Parametric} {Likelihood} with a {Focus} on {Bayesian} {Time} {Series} {Analysis}},
	shorttitle   = {Beyond {Whittle}},
	author       = {Kirch, Claudia and Edwards, Matthew C. and Meier, Alexander and Meyer, Renate},
	year         = 2019,
	month        = dec,
	journal      = {Bayesian Analysis},
	volume       = 14,
	number       = 4,
	doi          = {10.1214/18-BA1126},
	issn         = {1936-0975},
	url          = {https://projecteuclid.org/journals/bayesian-analysis/volume-14/issue-4/Beyond-Whittle--Nonparametric-Correction-of-a-Parametric-Likelihood-with/10.1214/18-BA1126.full},
	urldate      = {2025-09-17}
}

@misc{amaro-seoane_laser_2017,
	title        = {Laser {Interferometer} {Space} {Antenna}},
	author       = {Amaro-Seoane, Pau and others},
	year         = 2017,
	publisher    = {arXiv},
	doi          = {10.48550/ARXIV.1702.00786},
	url          = {https://arxiv.org/abs/1702.00786},
	urldate      = {2025-09-17},
	copyright    = {arXiv.org perpetual, non-exclusive license},
	note         = {arXiv:1702.00786, Version Number: 3}
}

@article{quang_nam_time-delay_2023,
	title        = {Time-delay interferometry noise transfer functions for {LISA}},
	author       = {Quang Nam, Dam and Martino, Joseph and Lemière, Yves and Petiteau, Antoine and Bayle, Jean-Baptiste and Hartwig, Olaf and Staab, Martin},
	year         = 2023,
	month        = oct,
	journal      = {Physical Review D},
	volume       = 108,
	number       = 8,
	pages        = {082004},
	doi          = {10.1103/PhysRevD.108.082004},
	issn         = {2470-0010, 2470-0029},
	url          = {https://link.aps.org/doi/10.1103/PhysRevD.108.082004},
	urldate      = {2025-09-17},
	language     = {en}
}

@misc{staab_pytdi_2025,
	title        = {{PyTDI}},
	author       = {Staab, Martin and Bayle, Jean-Baptiste and Hartwig, Olaf},
	year         = 2025,
	month        = sep,
	publisher    = {Zenodo},
	doi          = {10.5281/ZENODO.6351736},
	url          = {https://zenodo.org/doi/10.5281/zenodo.6351736},
    note         = {https://zenodo.org/doi/10.5281/zenodo.6351736},
	urldate      = {2025-09-17},
	copyright    = {BSD 3-Clause "New" or "Revised" License},
	note         = {Language: en}
}

@article{bayle_unified_2023,
	title        = {Unified model for the {LISA} measurements and instrument simulations},
	author       = {Bayle, Jean-Baptiste and Hartwig, Olaf},
	year         = 2023,
	month        = apr,
	journal      = {Physical Review D},
	volume       = 107,
	number       = 8,
	pages        = {083019},
	doi          = {10.1103/PhysRevD.107.083019},
	issn         = {2470-0010, 2470-0029},
	url          = {https://link.aps.org/doi/10.1103/PhysRevD.107.083019},
	urldate      = {2025-09-17},
	language     = {en}
}

@misc{bayle_lisa_2025,
	title        = {{LISA} {SGWB} {Dataset} (noise-4a)},
	author       = {Bayle, Jean-Baptiste},
	year         = 2025,
	month        = jun,
	publisher    = {Zenodo},
	doi          = {10.5281/ZENODO.15698080},
	url          = {https://zenodo.org/doi/10.5281/zenodo.15698080},
    note         = {https://zenodo.org/doi/10.5281/zenodo.15698080},
	urldate      = {2025-09-17},
	copyright    = {Creative Commons Attribution 4.0 International},
	language     = {en},
	collaborator = {{University of Glasgow}}
}

@article{roberts_examples_2009,
	title        = {Examples of {Adaptive} {MCMC}},
	author       = {Roberts, Gareth O. and Rosenthal, Jeffrey S.},
	year         = 2009,
	month        = jan,
	journal      = {Journal of Computational and Graphical Statistics},
	volume       = 18,
	number       = 2,
	pages        = {349--367},
	doi          = {10.1198/jcgs.2009.06134},
	issn         = {1061-8600, 1537-2715},
	url          = {http://www.tandfonline.com/doi/abs/10.1198/jcgs.2009.06134},
	urldate      = {2025-09-17},
	language     = {en}
}

@article{wood_p-splines_2017,
	title        = {P-splines with derivative based penalties and tensor product smoothing of unevenly distributed data},
	author       = {Wood, Simon N.},
	year         = 2017,
	month        = jul,
	journal      = {Statistics and Computing},
	volume       = 27,
	number       = 4,
	pages        = {985--989},
	doi          = {10.1007/s11222-016-9666-x},
	issn         = {0960-3174, 1573-1375},
	url          = {http://link.springer.com/10.1007/s11222-016-9666-x},
	urldate      = {2025-09-17},
	language     = {en}
}

@article{christensen_parameter_2022,
	title        = {Parameter estimation with gravitational waves},
	author       = {Christensen, Nelson and Meyer, Renate},
	year         = 2022,
	month        = apr,
	journal      = {Reviews of Modern Physics},
	volume       = 94,
	number       = 2,
	pages        = {025001},
	doi          = {10.1103/RevModPhys.94.025001},
	issn         = {0034-6861, 1539-0756},
	url          = {https://link.aps.org/doi/10.1103/RevModPhys.94.025001},
	urldate      = {2025-09-17},
	language     = {en}
}

@article{whittle_curve_1957,
	title        = {Curve and {Periodogram} {Smoothing}},
	author       = {Whittle, Peter},
	year         = 1957,
	journal      = {Journal of the Royal Statistical Society. Series B (Methodological)},
	volume       = 19,
	number       = 1,
	pages        = {38--63},
	issn         = {00359246},
	url          = {http://www.jstor.org/stable/2983994},
	urldate      = {2025-09-16},
	note         = {Publisher: [Royal Statistical Society, Oxford University Press]}
}

@article{Eilers_Marx_Durbán_2015,
  title        = {Twenty years of P-splines},
  author       = {Eilers, Paul H. C. and Marx, Brian D. and Durb{\'a}n, Maria},
  journal      = {SORT-Statistics and Operations Research Transactions},
  volume       = {39},
  number       = {2},
  pages        = {149--186},
  year         = {2015},
  month        = dec,
  url          = {https://raco.cat/index.php/SORT/article/view/302258}
}

@article{maturana-russel_bayesian_2021,
	title        = {Bayesian spectral density estimation using {P}-splines with quantile-based knot placement},
	author       = {Maturana-Russel, Patricio and Meyer, Renate},
	year         = 2021,
	month        = sep,
	journal      = {Computational Statistics},
	volume       = 36,
	number       = 3,
	pages        = {2055--2077},
	doi          = {10.1007/s00180-021-01066-7},
	issn         = {0943-4062, 1613-9658},
	url          = {https://link.springer.com/10.1007/s00180-021-01066-7},
	urldate      = {2025-09-17},
	language     = {en}
}

@article{baghi_uncovering_2023,
	title        = {Uncovering gravitational-wave backgrounds from noises of unknown shape with {LISA}},
	author       = {Baghi, Quentin and Karnesis, Nikolaos and Bayle, Jean-Baptiste and Besançon, Marc and Inchauspé, Henri},
	year         = 2023,
	month        = apr,
	journal      = {Journal of Cosmology and Astroparticle Physics},
	volume       = 2023,
	number       = {04},
	pages        = {066},
	doi          = {10.1088/1475-7516/2023/04/066},
	issn         = {1475-7516},
	url          = {https://iopscience.iop.org/article/10.1088/1475-7516/2023/04/066},
	urldate      = {2025-09-17}
}

@article{flauger_improved_2021,
	title        = {Improved reconstruction of a stochastic gravitational wave background with {LISA}},
	author       = {Flauger, Raphael and Karnesis, Nikolaos and Nardini, Germano and Pieroni, Mauro and Ricciardone, Angelo and Torrado, Jesús},
	year         = 2021,
	month        = jan,
	journal      = {Journal of Cosmology and Astroparticle Physics},
	volume       = 2021,
	number       = {01},
	pages        = {059--059},
	doi          = {10.1088/1475-7516/2021/01/059},
	issn         = {1475-7516},
	url          = {https://iopscience.iop.org/article/10.1088/1475-7516/2021/01/059},
	urldate      = {2025-09-17},
	copyright    = {http://iopscience.iop.org/info/page/text-and-data-mining}
}

@article{cornish_bayeswave_2015,
	title        = {Bayeswave: {Bayesian} inference for gravitational wave bursts and instrument glitches},
	shorttitle   = {Bayeswave},
	author       = {Cornish, Neil J. and Littenberg, Tyson B.},
	year         = 2015,
	month        = jul,
	journal      = {Classical and Quantum Gravity},
	volume       = 32,
	number       = 13,
	pages        = 135012,
	doi          = {10.1088/0264-9381/32/13/135012},
	issn         = {0264-9381, 1361-6382},
	url          = {https://iopscience.iop.org/article/10.1088/0264-9381/32/13/135012},
	urldate      = {2025-09-17}
}

@article{biscoveanu_quantifying_2020,
	title        = {Quantifying the effect of power spectral density uncertainty on gravitational-wave parameter estimation for compact binary sources},
	author       = {Biscoveanu, Sylvia and Haster, Carl-Johan and Vitale, Salvatore and Davies, Jonathan},
	year         = 2020,
	month        = jul,
	journal      = {Physical Review D},
	volume       = 102,
	number       = 2,
	pages        = {023008},
	doi          = {10.1103/PhysRevD.102.023008},
	issn         = {2470-0010, 2470-0029},
	url          = {https://link.aps.org/doi/10.1103/PhysRevD.102.023008},
	urldate      = {2025-09-17},
	language     = {en}
}

@article{punturo_einstein_2010,
	title        = {The {Einstein} {Telescope}: a third-generation gravitational wave observatory},
	shorttitle   = {The {Einstein} {Telescope}},
	author       = {Punturo, M and Abernathy, M and Acernese, F and Allen, B and Andersson, N and Arun, K and Barone, F and Barr, B and Barsuglia, M and Beker, M and others},
	year         = 2010,
	month        = oct,
	journal      = {Classical and Quantum Gravity},
	volume       = 27,
	number       = 19,
	pages        = 194002,
	doi          = {10.1088/0264-9381/27/19/194002},
	issn         = {0264-9381, 1361-6382},
	url          = {https://iopscience.iop.org/article/10.1088/0264-9381/27/19/194002},
	urldate      = {2025-09-17}
}

@article{abbott_guide_2020,
	title        = {A guide to {LIGO}–{Virgo} detector noise and extraction of transient gravitational-wave signals},
	author       = {Abbott, B P and Abbott, R and Abbott, T D and Abraham, S and Acernese, F and Ackley, K and Adams, C and Adya, V B and Affeldt, C and Agathos, M and others},
	year         = 2020,
	month        = mar,
	journal      = {Classical and Quantum Gravity},
	volume       = 37,
	number       = 5,
	pages        = {055002},
	doi          = {10.1088/1361-6382/ab685e},
	issn         = {0264-9381, 1361-6382},
	url          = {https://iopscience.iop.org/article/10.1088/1361-6382/ab685e},
	urldate      = {2025-09-17}
}

@article{welch_use_1967,
	title        = {The use of fast {Fourier} transform for the estimation of power spectra: {A} method based on time averaging over short, modified periodograms},
	shorttitle   = {The use of fast {Fourier} transform for the estimation of power spectra},
	author       = {Welch, Peter D.},
	year         = 1967,
	month        = jun,
	journal      = {IEEE Transactions on Audio and Electroacoustics},
	volume       = 15,
	number       = 2,
	pages        = {70--73},
	doi          = {10.1109/TAU.1967.1161901},
	issn         = {0018-9278},
	url          = {http://ieeexplore.ieee.org/document/1161901/},
	urldate      = {2025-09-17},
	copyright    = {https://ieeexplore.ieee.org/Xplorehelp/downloads/license-information/IEEE.html},
	language     = {en}
}

@article{littenberg_prototype_2023,
	title        = {Prototype global analysis of {LISA} data with multiple source types},
	author       = {Littenberg, Tyson B. and Cornish, Neil J.},
	year         = 2023,
	month        = mar,
	journal      = {Physical Review D},
	volume       = 107,
	number       = 6,
	pages        = {063004},
	doi          = {10.1103/PhysRevD.107.063004},
	issn         = {2470-0010, 2470-0029},
	url          = {https://link.aps.org/doi/10.1103/PhysRevD.107.063004},
	urldate      = {2025-09-17},
	language     = {en}
}

@article{Umstatter:2005jd,
    author = "Umst{\"a}tter, Richard and Christensen, Nelson and Hendry, Martin and Meyer, Renate and Simha, Vimal and Veitch, John and Vigeland, Sarah and Woan, Graham",
    title = "{Bayesian modeling of source confusion in LISA data}",
    eprint = "gr-qc/0506055",
    archivePrefix = "arXiv",
    doi = "10.1103/PhysRevD.72.022001",
    journal = "Physical Review D",
    volume = "72",
    pages = "022001",
    year = "2005"
}

@article{Green:1995mxx,
    author = "Green, Peter J.",
    title = "{Reversible jump Markov chain Monte Carlo computation and Bayesian model determination}",
    doi = "10.1093/biomet/82.4.711",
    journal = "Biometrika",
    volume = "82",
    number = "4",
    pages = "711--732",
    year = "1995"
}

@article{Christensen_2018,
	doi = {10.1088/1361-6633/aae6b5},
	url = {https://doi.org/10.1088/1361-6633/aae6b5},
	year = 2018,
	month = {11},
	publisher = {{IOP} Publishing},
	volume = {82},
	number = {1},
	pages = {016903},
	author = {Nelson Christensen},
	title = {Stochastic gravitational wave backgrounds},
	journal = {Reports on Progress in Physics}
}

@article{Deng:2025wgk,
    author = "Deng, Senwen and Babak, Stanislav and Le Jeune, Maude and Marsat, Sylvain and Plagnol, {\'E}ric and Sartirana, Andrea",
    title = "{Modular global-fit pipeline for LISA data analysis}",
    eprint = "2501.10277",
    archivePrefix = "arXiv",
    primaryClass = "gr-qc",
    doi = "10.1103/PhysRevD.111.103014",
    journal = "Physical Review D",
    volume = "111",
    number = "10",
    pages = "103014",
    year = "2025"
}

@article{Katz:2024oqg,
    author = "Katz, Michael L. and Karnesis, Nikolaos and Korsakova, Natalia and Gair, Jonathan R. and Stergioulas, Nikolaos",
    title = "{Efficient GPU-accelerated multisource global fit pipeline for LISA data analysis}",
    eprint = "2405.04690",
    archivePrefix = "arXiv",
    primaryClass = "gr-qc",
    doi = "10.1103/PhysRevD.111.024060",
    journal = "Physical Review D",
    volume = "111",
    number = "2",
    pages = "024060",
    year = "2025"
}

@article{Littenberg:2023xpl,
    author = "Littenberg, Tyson B. and Cornish, Neil J.",
    title = "{Prototype global analysis of LISA data with multiple source types}",
    eprint = "2301.03673",
    archivePrefix = "arXiv",
    primaryClass = "gr-qc",
    doi = "10.1103/PhysRevD.107.063004",
    journal = "Physical Review D",
    volume = "107",
    number = "6",
    pages = "063004",
    year = "2023"
}

@article{Boileau:2021sni,
    author = "Boileau, Guillaume and Lamberts, Astrid and Christensen, Nelson and Cornish, Neil J. and Meyer, Renate",
    title = "{Spectral separation of the stochastic gravitational-wave background for LISA in the context of a modulated Galactic foreground}",
    eprint = "2105.04283",
    archivePrefix = "arXiv",
    primaryClass = "gr-qc",
    doi = "10.1093/mnras/stab2575",
    journal = "Monthly Notices of the Royal Astronomical Society",
    volume = "508",
    number = "1",
    pages = "803--826",
    year = "2021",
    note = "[Erratum: Mon.Not.Roy.Astron.Soc. 508, 5554--5555 (2021)]"
}

@article{Janssens:2022xmo,
    author = "Janssens, Kamiel and Boileau, Guillaume and Christensen, Nelson and Badaracco, Francesca and van Remortel, Nick",
    title = "{Impact of correlated seismic and correlated Newtonian noise on the Einstein Telescope}",
    eprint = "2206.06809",
    archivePrefix = "arXiv",
    primaryClass = "astro-ph.IM",
    doi = "10.1103/PhysRevD.106.042008",
    journal = "Physical Review D",
    volume = "106",
    number = "4",
    pages = "042008",
    year = "2022"
}

@book{Ruppert2003,
  title={Semiparametric regression},
  author={Ruppert, David and Wand, Matt P. and Carroll, Raymond J.},
  number={12},
  year={2003},
  publisher={Cambridge university press}
}

@misc{scikit-fda-repo,
  author = {The scikit-fda developers},
  doi = {10.5281/zenodo.3468127},
  month = "feb",
  title = {scikit-fda: Functional Data Analysis in Python},
  url = {https://github.com/GAA-UAM/scikit-fda},
  year = {2024}
}

@article{scikit-fda-paper,
  author = {Ramos-Carreño, Carlos and Torrecilla, José L. and Carbajo Berrocal, Miguel and Marcos Manchón, Pablo and Suárez, Alberto},
  doi = {10.18637/jss.v109.i02},
  journal = {Journal of Statistical Software},
  month = "may",
  number = {2},
  pages = {1--37},
  title = {{scikit-fda: A Python Package for Functional Data Analysis}},
  url = {https://www.jstatsoft.org/article/view/v109i02},
  volume = {109},
  year = {2024}
}

@misc{aimen_bayesian_2025,
	title        ={Dataset for {B}ayesian {PSD} estimation for {LISA} noise based on {P}-splines with a parametric boost},
	author       = {Aimen, Nazeela and Maturana-Russel, Patricio and Vajpeyi, Avi and Christensen, Nelson and Meyer, Renate},
	year         = 2025,
	month        = sep,
	publisher    = {Zenodo},
	doi          = {10.5281/ZENODO.17116773},
	url          = {https://zenodo.org/doi/10.5281/zenodo.17116773},
    note         = {https://zenodo.org/doi/10.5281/zenodo.17116773},
	urldate      = {2025-12-10},
	copyright    = {Creative Commons Attribution 4.0 International},
	language     = {en},
}

\end{document}